\documentclass{aa}

\def\alt{\hbox{\raise.5ex\hbox{$<$}
\kern-1.1em\lower.5ex\hbox{$\sim$}}}
\def\agt{\hbox{\raise.5ex\hbox{$>$}
\kern-1.1em\lower.5ex\hbox{$\sim$}}}

\begin{document}

\thesaurus{02(12.03.3; 08.19.4)}

\title{Do we really see a cosmological constant in the supernovae data ?} 

\author{Marie-No\"elle C\'EL\'ERIER}

\institute{D\'epartement d'Astrophysique Relativiste et de 
Cosmologie, Observatoire de Paris-Meudon, \\
5 place Jules Janssen, 92195 Meudon C\'edex, France}

\mail{celerier@obspm.fr}

\date{Received 7 July 1999 / Accepted 29 September 1999}

\maketitle
\markboth{M.N. C\'el\'erier: Do we really see a Cosmological Constant in
the Supernovae data ?}
{M.N. C\'el\'erier: Do we really see a Cosmological Constant in
the Supernovae data ?}

\begin{abstract}
The magnitude-redshift relation is one of the tools for a direct 
observational approach to cosmology. The discovery of high redshift 
Type Ia supernovae (SNIa) and their use as ``standard candles'' has 
resurrected interest in this approach. Recently collected  data 
have been used to address the problem of measuring the cosmological 
parameters of the universe. Analysed in the framework of  
homogeneous models, they have yielded, as a primary result, a strictly 
positive cosmological constant. However, a straight reading of the 
published measurements, conducted with no a priori idea of which 
model would best describe our universe at least 
up to redshifts $z\sim 1$, does not exclude the possibility 
of ruling out the Cosmological Principle  - and cosmological constant - 
hypotheses. It is therefore shown here how the large scale homogeneity 
of this part of the universe can be tested on our past light cone, 
using the magnitude-redshift relation, provided sufficiently accurate 
data from sources at redshifts approaching $z=1$ would be available. 
An example of an inhomogeneous model 
with zero cosmological constant reproducing the current observations 
is given. The presently published SNIa data can thus be interpreted as 
implying either a strictly positive cosmological constant in a 
homogeneous universe or large scale inhomogeneity with no constraint 
on $\Lambda$. An increase in the number and measurement accuracy of the 
candidate ``standard candles'' at very high redshift is therefore 
urgently needed, for progress in both fundamental issues of the 
Cosmological Principle and of the cosmological constant.

\end{abstract}

\bigskip
\section{Introduction}

The use of astronomical observations to directly determine the 
space-time geometry of our universe is a long time proposed instrument 
(\cite{sandage,kristian,ellis85}). First presented as an ideal program 
which could be realised ``in principle'', it has of late become a new 
field in cosmology, due to recent tremendous progress in 
instrumental technology. \\

The magnitude-redshift relation is one of the tools for a direct 
observational approach. The discovery of high redshift ($z\sim 0.5-1$) 
Type Ia supernovae (SNIa), and of their potential use as ``standard 
candles'' (\cite{phillips,riess95,hamuy95,hamuy96a,hamuy96b}, 
Perlmutter et al. \cite{perlmutter97,tripp,jha}) has resurrected 
interest in this 
approach. Recently collected  data, from ongoing systematic searches, 
have been used to address the problem of measuring the cosmological 
parameters of the universe. Analysed in the framework of homogeneous 
models, they have yielded, as a primary result, a strictly positive 
cosmological constant, many orders of magnitude smaller than the energy 
of the vacuum expected in standard particle physics models, as 
proposed by Riess et al. (\cite{riess}), hereafter refered to as R98, 
and by Perlmutter et al. (\cite{perlmutter99}), hereafter refered to 
as P99. \\

A non zero cosmological constant was long ago proposed as a 
possible interpretation of magnitude-redshift observations of cluster 
galaxies and quasars by Solheim (\cite{solheim}). But the accuracy of 
the data was not then sufficient to probe this hypothesis. \\

Anything that contributes to the energy density of the vacuum acts 
just like a cosmological constant. Analysed in the framework of 
Friedmann models of the universe, this can be viewed as an argument 
in favor of a zero cosmological constant, since some unknown symmetry 
of particle physics theory could presumably cancel  the vacuum energy 
density (see e.g., \cite{weinberg}). If the value of the cosmological 
constant was confirmed to be in the range favored by the SNIa 
announced results, it would be necessary to explain how it is so small, 
yet non zero. As such a result would have a revolutionary impact on our 
understanding of the fundamental laws of physics, it is of tremendous 
importance to check the different available interpretations.\\

Provided every other source of potential bias or systematic 
uncertainties has been correctly taken into account, one of the 
most vulnerable points of the data collecting procedure is the 
difference in the absolute magnitude of the supernovae due to an 
evolution of their progenitors. This point is discussed at length 
in R98 and P99 who conclude that their results cannot be markedly 
affected by this effect. Other authors (\cite{drell,dominguez}) 
claim contrary evidence that high and low redshift supernovae 
observed so far, substantially differ from one another, and thus 
contest their use as ``standard candles''. A progressive dimming of 
the SNIa by intergalactic dust has also been proposed as a possible 
systematic effect that could mimic the behaviour of cosmic acceleration, 
and thus allow cosmologies without a cosmological constant 
(\cite{aguirre}). This important problem is 
not treated in the present work, where the working assumption is made 
that the SNIa data actually measure the magnitude-redshift relation up 
to the precisions claimed in R98 and P99. \\

The theoretical interpretation is another matter 
which has to be dealt with carefully to avoid a priori assertions which 
would lead to incorrect results. \\

P98 discuss the possibility that the magnitude-redshift 
relation they find from the analysis of their data is due, not to a 
cosmological constant, but to an evolving field of unknown nature that 
contributes to the total energy density of the universe (see e.g., 
\cite{steinhardt,caldwell,garnavich}). This entity, which can have 
an equation-of-state ratio different from that of the cosmological 
constant, would lead to a different expansion history. \\

It is proposed here to focus on a less exotic alternative, namely the 
possibility of large scale inhomogeneity of the 
part of the universe which can be probed with SNIa measurements. \\

Large scale spatial homogeneity of the universe is a belief most 
commonly shared by current cosmologists. It procceeds from a 
hypothesis brought to the status of Cosmological Principle by Einstein 
(\cite{einstein}). Its justification is based on two arguments: \\
\begin{enumerate}
\item The isotropy, or quasi-isotropy, of the temperature of the Cosmic 
Microwave Background Radiation (CMBR) around us. \\
\item The Copernican assumption that, as our location must not be special, 
the isotropy we observe must be observed identically from any point of 
the universe. \\
\end{enumerate}

Since a matter distribution which is seen isotropic from everywhere 
implies homogeneity, the Cosmological Principle follows. However, of 
the two above arguments, only the first is observation 
grounded. The second, purely philosophical, has never been  
verified, and cannot thus constitute acceptable evidence. \\

As will be shown in Sect. 2, the claim for a strictly positive 
cosmological constant from SNIa data procceeds from an a priori 
homogeneity assumption. This, and the central role played in the 
whole cosmological field by the Cosmological Principle, is sufficient 
motivation to examine to which extent this Principle can be 
observationally tested. \\

As has been stressed by Ellis (\cite{ellis79}), the situation is 
completly different within and outside our past light cone. For models 
exhibiting particle horizons, there are regions of the universe, 
outside our past light cone, from which we cannot receive any 
information at the present time. There is thus no way we can 
observationally verify the spatial homogeneity of such far out 
regions. \\

The possibility of direct verification on our past light cone has 
been studied by Partovi and Mashhoon (\cite{partovi}). These authors 
have explored the extent to which it is possible to distinguish 
between large scale homogeneous and inhomogeneous spherically 
symmetrical models, using magnitude-redshift data. Owing to the then 
state of observational technique, they concluded this distinction 
could not be performed. \\

Analyses of galaxy redshift surveys have recently received increased 
attention. A controversy over whether the universe is smooth on large 
scale (\cite{guzzo,cappi,martinezetal,scaramella,wu}) or presents an 
unbounded fractal hierarchy (\cite{sylos1996,sylos1998}) has developed 
and waits for the next generation of wider and deeper galaxy catalogues 
which may provide a more conclusive answer (\cite{martinez}). \\

The supernovae data have already been investigated for what they can 
tell about (in)homogeneity. Kim et al. (\cite{kim}) used the first seven 
SNIa discovered by the Supernova Cosmology Project at $0.35 < z < 0.65$ 
and compared them to a nearby sample at $z \leq 0.1$ to declare the 
ruling out of the hypothesis of a locally underdense bubble. The nearby 
sample has also been examined for evidence of a local ``Hubble Bubble'' 
by Zehavi et al. (\cite{zehavi}). The marginal signal identified by 
these authors is very sensitive to the cosmological model retained. Its 
value has been estimated for an Einstein-de Sitter universe, but would 
be less significant, and could be considered as included in the limits 
calculated by Kim et al., for the type of models proposed by R98 and 
P99. \\

At another redshift extreme, Starkman et al. (\cite{starkman}) have 
shown that the segment of the universe sampled by the current supernovae 
data is not large enough to determine the overall properties of the 
expansion. But these authors only consider the Einstein-de Sitter 
homogeneous case to complete their caculations. \\

It will however be shown, in Sect.2, that in the near future, we would 
in principle, be able to probe homogeneity on scales up to, at least 
$z=1$. In principle here means provided the evolution effect comes 
under control, as well as any other bias or systematic effect. But, as 
long as the relevance of the homogeneous models used by R98 and P99 as 
a framework for their data analyses is not verified, any claim to   
cosmological parameter measurements remains premature. Furthermore, 
a straight reading of the presently published results does not exclude 
a ruling out of the homogeneity hypothesis. It is moreover 
shown, in Sect. 3 and 4, that large scale 
inhomogeneity can mimic a cosmological constant in an homogeneous 
universe up to the precision achieved by current measurements.

\section{Magnitude-redshift relation and homogeneous models}

The luminosity distance $D_L$ of a source is defined as the distance 
from which the radiating body, if motionless in an Euclidean space, 
would produce an energy flux equal to the one measured by the observer. 
It thus verifies

\begin{equation}
{l=}{L\over {4\pi D_L^2}} , \label{eq:1}
\end{equation}
$L$ being the absolute luminosity, i.e. the luminosity in the rest 
frame of the source, and $l$ the measured bolometric flux, i.e. 
integrated over all frequencies by the observer. \\

In Friedmann-Lema\^itre-Robertson-Walker (FLRW) models, the distance 
measure at redshift $z$ is a function of $z$ and of the parameters of 
the model (\cite{carroll})

\begin{eqnarray}
D_L={c(1+z)\over {H_0\sqrt{|\kappa |}}} \,\, {\cal S}&\biggl(&\sqrt
{|\kappa |}\int^z_0[(1+z')^2(1+\Omega _M z') \nonumber \\
&&-z'(2+z')\Omega _\Lambda]^{-{1\over 2}}
dz'\biggr) , \label{eq:2}
\end{eqnarray}
$H_0$ being the current Hubble constant, $\Omega _M$, the mass density 
parameter, and, $\Omega _\Lambda$ being defined as
\begin{eqnarray}
\Omega _\Lambda \equiv {\Lambda\over {3H_0^2}} , \nonumber
\nonumber
\end{eqnarray}
where $\Lambda$ is the cosmological constant. \\
and with \\

\begin{tabular}{cll}
for $\Omega _M + \Omega _\Lambda > 1$ & ${\cal S}=\sin$ & and 
$\kappa = 1-\Omega _M-\Omega _\Lambda$ \\
for $\Omega _M + \Omega _\Lambda < 1$ & ${\cal S}=\sinh$ & and 
$\kappa = 1-\Omega _M-\Omega _\Lambda$ \\
for $\Omega _M + \Omega _\Lambda = 1$ & ${\cal S}=$ I & and 
$\kappa = 1$ . \\
\end{tabular}
\\

The apparent bolometric magnitude $m$ of a standard candle of absolute 
bolometric magnitude $M$, at a given redshift $z$, is thus also a 
function of $z$ and of the parameters of the model. Following 
Perlmutter et al. (\cite{perlmutter97}), hereafter referred to as P97, 
it can be written, in units of megaparsecs, as
\begin{eqnarray}
m&=&M+5 \log D_L(z;\Omega _M,\Omega _\Lambda, H_0)+25 
\nonumber \\
&\equiv & {\cal M}+5 \log {\cal D}_L(z;\Omega _M,\Omega _\Lambda) .
\label{eq:3}
\end{eqnarray}

The magnitude ``zero-point'' ${\cal M}\equiv M-5 \log H_0+25$ can be 
measured from the apparent magnitude and redshift of low-redshift 
examples of the standard candles, without knowing $H_0$. Furthermore, 
${\cal D}_L(z;\Omega _M,\Omega _\Lambda)\equiv H_0 D_L(z;\Omega _M,
\Omega _\Lambda, H_0)$  depends on $\Omega _M$ and 
$\Omega _\Lambda$ with different functions of redshift. A priori 
assuming that FLRW models are valid to describe the observed universe, 
R98 and P99 thus proceed as follows to determine $\Omega _M$ and 
$\Omega _\Lambda$. \\

A set of apparent magnitude and redshift measurements for low-redshift 
($0.004 \ \alt \ z\  \alt \ 0.1$) SNIa is used to calibrate 
Eq.(\ref{eq:3}), and another set of such measurements for high-redshift 
($0.16 \ \alt \ z \ \alt \ 0.97$) is used to determine the best fit 
values 
of $\Omega _M$ and $\Omega _\Lambda$\footnote{The methods used by the 
two survey teams, the stretch factor method for P99 and the multicolor 
light curve shape and template fitting methods for R98, differ in their 
conceptions and byproducts: P99 calibrate ${\cal M}$ and obtain direct 
fitting of $\Omega _M$ and $\Omega _\Lambda$; R98 calibrate $M$, which 
yields an estimate for $H_0$. However, the following remarks apply to 
each of them.}. The results of these measurements 
are plotted in Fig. 4 and 5 of R98 and Fig. 1 and 2 of P99. The 
magnitude-redshift relation from the data and the theoretical curves 
obtained for different values of the cosmological parameters are then 
compared and best-fit confidence regions in the parameter 
space are plotted, see Fig. 6 and 7 of R98 and Fig. 7 of P99.\\

It is convenient at this stage to make the following remarks. \\

If one considers any cosmological model for which the luminosity 
distance $D_L$ is a function of the redshift $z$ and of the other 
cosmological parameters of the model, and if this function is 
Taylor expandable near the observer, i.e. around $z=0$, the analysis 
of observational data at $z<1$, in the framework of this model, can 
legitimately use the Taylor expansion

\begin{eqnarray}
&&D_L(z;cp)=\left(dD_L\over dz\right)_{z=0}z \, + \, {1\over 2}
\left(d^2D_L\over dz^2\right)_{z=0}z^2  \nonumber \\
&&+ \, {1\over 6}\left(d^3D_L \over dz^3\right)_{z=0}z^3 \, + \, 
{1\over 24}\left(d^4D_L\over dz^4\right)_{z=0}z^4 + {\cal O}(z^5) , 
\label{eq:4}
\end{eqnarray}
as, by definition of luminosity distances, $D_L(z=0)=0$. \\
Here, $cp$ denotes the set of cosmological parameters, pertaining to 
the given model, which can be 
either constants, as in FLRW models, or functions of $z$, as in the 
example presented in Sect. 3 and 4. \\

Luminosity distance measurements of sources at different redshifts 
$z<1$ yield values for the different coefficients in the above 
expansion. Going to higher redshifts amounts to measuring the 
coefficients of higher power of $z$. For very low redshifts, the 
leading term is first order; for intermediate redshifts, second 
order. Then third order terms provide significant contributions. 
For redshifts approaching unity, higher order terms can no longer be 
neglected. \\

Therefore, for cosmological models with very high (or infinite) 
number of free parameters, such data only provide 
constraints upon the values of the parameters near the observer. 
But for cosmological models with few parameters, giving independent 
contributions to each coefficient in the expansion, the method not 
only provides a way to evaluate the parameters, but, in most cases, 
to test the validity of the model itself. \\

For FLRW models, precisely, one obtains from Eq.(\ref{eq:2})
\begin{eqnarray}
D^{(1)}_L\equiv \left(dD_L\over dz\right)_{z=0}&=&{c\over H_0} , 
\label{eq:5} \\
D^{(2)}_L\equiv {1\over 2}\left(d^2D_L\over dz^2\right)_{z=0}&=&
{c\over {4 H_0}}(2-\Omega _M+2 \, \Omega _\Lambda) , \label{eq:6} \\
D^{(3)}_L\equiv {1\over 6}\left(d^3D_L\over dz^3\right)_{z=0}&=&
{c\over {8 H_0}}(-2 \, \Omega _M-4 \, \Omega _\Lambda-4 \, \Omega _M  
\Omega _\Lambda \nonumber \\
&&+\Omega^2_M+4 \, \Omega^2_\Lambda) , \label{eq:7} \\
D^{(4)}_L\equiv {1\over 24}\left(d^4D_L\over dz^4\right)_{z=0}&=&
{5 c\over {72 H_0}}\bigl(8 \, \Omega _\Lambda+4 \, \Omega _M  
\Omega _\Lambda+2 \, \Omega^2_M \nonumber \\
-16 \, \Omega^2_\Lambda -12 \, \Omega _M  \Omega^2_\Lambda 
&+&6 \, \Omega^2_M \Omega _\Lambda-\Omega^3_M+8 \, \Omega^3_\Lambda\bigr) 
. \label{eq:8} \\
\nonumber
\end{eqnarray}

These coefficients are independent functions of the three parameters 
of this class of  models, $H_0, \Omega _M$ and $\Omega _\Lambda$. 
With the method retained by P99 and described above, $H_0$ is hidden 
in the magnitude ``zero-point'' $\cal M$. The coefficients 
${\cal D}^{(i)}_L$ of the expansion of ${\cal D}_L$ are thus functions 
of $\Omega _M$ and $\Omega _\Lambda$ alone. However, the same 
following remarks, with $D$ replacing ${\cal D}$, 
apply to the analysis and results of R98. \\

In standard models, i.e. for $\Omega _\Lambda=0$
\begin{equation}
{\cal D}^{(2)}_L={c\over 4 }(2-\Omega _M) . 
\label{eq:9}
\end{equation}

If the analysis of the measurements gives ${\cal D}^{(1)}_L<2\ 
{\cal D}^{(2)}_L$, it implies $\Omega _M<0$, which is physically 
irrelevant. The model is thus ruled out. This is what happens with the 
SNIa data, and what induces R98 and P99 to postulate a strictly 
positive cosmological constant, to counteract the $-\Omega _M$ 
term. \\

To test FLRW models with $\Omega _\Lambda\neq 0$, one has to go at 
least to the third order to have a chance to obtain a result. This 
seems to be the order currently reached by the SNIa surveys. This 
last assertion rests on the two following remarks: \\
\begin{enumerate}
\item  As stressed in P97, R98 and P99, the well-constrained linear 
combination of $\Omega _M$ and $\Omega _\Lambda$ obtained from the 
data is not parallel to any contour of constant current 
``deceleration''\footnote{A better name would be 
``acceleration'' parameter, as $q_0<0$ implies an accelerated 
expansion for a homogeneous universe.} parameter 
$q_0=\Omega _M \hbox {/2}-
\Omega _\Lambda$. As ${\cal D}^{(2)}_L={\cal D}^{(1)}_L(1-q_0)\hbox 
{/2}$, this implies that higher order terms effectively 
contribute. \\
\item The contribution of the fourth and higher order terms is 
negligeable. One can easily verify that, for the higher redshifts 
reached by the surveys, the contribution of the fourth order term does 
not overcome the measurement uncertainties, of the order of 5 
to 10\% (see Fig. 4 and 5 of R98 and Fig. 1 and 2 of P99). \\
\end{enumerate}

A ruling out of FLRW models with non zero cosmological constant, 
at the third order level, i.e. due to a negative value for 
$\Omega _M$, would occur provided (see Eq.(\ref{eq:5}) to 
(\ref{eq:7}))
\begin{equation}
1-{{\cal D}^{(3)}_L\over {\cal D}^{(1)}_L}-3 \, {{\cal D}^{(2)}_L\over 
{\cal D}^{(1)}_L}+2 \left ({\cal D}^{(2)}_L\over {\cal D}^{(1)}_L
\right)^2<0 . 
\label{eq:10}
\end{equation}

This cannot be excluded by the results of R98 and P99. It 
corresponds to the left part of the truncated best-fit confidence 
ellipsoidal regions in the $\Omega _M-\Omega _\Lambda$ plane of R98 
Fig. 6 and 7 and P99 Fig. 7, and to the upper part of the error bars 
for the higher redshifts data in R98 Fig. 4 and 5 and P99 Fig. 1 and 
2. \\

P99 propose an approximation of their results, which they write as
\begin{equation}
0.8\, \Omega _M-0.6\, \Omega _\Lambda \approx - \, 0.2\, \pm \, 0.1 , 
\label{eq:11}
\end{equation}
which corresponds, in fact, to:
\begin{eqnarray}
2\left({\cal D}^{(2)}_L\over{\cal D}^{(1)}_L\right)^2-\, 4.2 \, {{\cal 
D}^{(2)}_L\over {\cal D}^{(1)}_L}-{{\cal D}^{(3)}_L\over {\cal D}^
{(1)}_L}\approx -1.8\,  \pm \, 0.1  . \label{eq:12} \\
\nonumber
\end{eqnarray}

It must be here stressed that the results published as primary by P99 
and R98, under the form of best-fit confidence regions in the 
$\Omega _M-\Omega _\Lambda$ plane, proceed from a Bayesian data 
analysis, for which a prior probability distribution accounting for the 
physically (in Friedmann cosmology) allowed part of parameter space is 
assumed. Results as given in the form of Eq.(\ref{eq:11}) are thus 
distorted by an a priori homogeneity assumption, which would have to be 
discarded for the completion of the test here proposed. \\

It may however occur that future, more accurate measurements yield 
values for the ${\cal D}^{(i)}_L$s verifying Eq.(\ref{eq:10}) with the 
$>$ sign, which would correspond to a physically consistent positive 
$\Omega _M$. In this 
case, the FLRW models will have to be tested to the fourth order, 
i.e. with sources at redshifts nearer $z=1$. A final test for the 
homogeneity hypothesis on our past light cone would be a check of the 
necessary condition, obtained from Eq.(\ref{eq:5}) to (\ref{eq:8}), and 
which can be written as
\begin{eqnarray}
{\cal D}^{(4)}_L&=&- \, {10\over 9} \, {\cal D}^{(2)}_L+{10\over 3} \, 
{{\cal D}^{(2)2}_L\over {\cal D}^{(1)}_L}-{20\over 9} \, 
{{\cal D}^{(2)3}_L\over {\cal D}^{(1)2}_L}-{20\over 9} \, 
{\cal D}^{(3)}_L \nonumber \\
&&+{10\over 3} \, {{\cal D}^
{(2)}_L{\cal D}^{(3)}_L\over {\cal D}^{(1)}_L} . \label{eq:13} \\
\nonumber
\end{eqnarray} 

The above described method only applies to data issued from $z<1$ 
supernovae. If the ongoing surveys were to discover more distant sources, 
up to redshifts higher than unity, the Taylor expansion would no 
longer be valid. \\

In practice, one will have to consider the Hubble diagram for the 
largest sample of accurately measured standard candles at every 
available scale of redshift from the lowest to the highest (e.g. 
Fig. 4 and 5 of R98 and Fig. 1 and 2 of P99). The FLRW theoretical 
diagram best fitting the data at intermediate redshift ($z\approx 
0.6-0.7$ for measurements with 5\% accuracy, $z\approx 0.8$ for 10\% 
accuracy) will be retained as candidate. If it corresponds to a 
negative value for $\Omega _M$, the assumption of large scale 
homogeneity for the observed part of the universe will have to be 
discarded, 
and no higher redshift data will be needed, at least to deal with this 
issue. If it corresponds to a positive $\Omega _M$, the diagram will 
have to be extended to higher redshift data. If these data confirm 
the best fit of this candidate model up to $z\approx 0.8-0.9$ (for 
measurements at some 5-10\% accuracy), the homogeneity hypothesis (and 
the corresponding values of the model parameters) would receive a 
robust support (provided no unlikely fine tuning of another model of 
universe parameters). \\

Another practical way of analysing the data is given by the very clever 
method described by Goobar and Perlmutter (\cite{goobar}), hereafter 
refered to as 
GP. An application of this method to  probe large scale homogeneity of 
the observed universe would, in principle, imply the measurement of 
the apparent magnitude and redshift of only three intermediate and 
high redshift standard candles. In practice, more would certainly 
be needed to smooth out observational uncertainties. The method applies 
to sources with any redshift values and runs as follows. \\

Using Eq.(\ref{eq:2}) and (\ref{eq:3}), one can predict the apparent 
magnitude of a source measured at a given redshift, in a peculiar 
FLRW model (i.e. for a given pair of values for $\Omega _M$ and 
$\Omega _\Lambda$). Fig. 1 of GP shows the contours of constant apparent 
magnitude on the $\Omega _\Lambda$-versus-$\Omega _M$ plane for two 
sufficiently different redshifts. When an actual apparent 
measurement of a source at a given redshift is made, the candidate FLRW 
model selected is the one with values for $\Omega _M$ and $\Omega 
_\Lambda$ narrowed to a single contour line. Since one can assume some 
uncertainty in the measurements, the allowed ranges of $\Omega _M$ 
and $\Omega _\Lambda$ are given by a strip between two contour lines. 
Two such measurements for sources at different redshift can define two 
strips that cross in a more narrowly constrained ``allowed'' region,
shown as a dashed rhombus in Fig. 1 of GP. Now, if a third measurement 
of a source at a redshift sufficiently different from the two others is 
made, a third strip between two contour lines is selected. If this 
strip clearly crosses the previously drawn rhombus (and if the 
measurement uncertainties are kept within acceptable values), it 
provides support for the homogeneity hypothesis, with the above 
pointed out reserve. In case this strip clearly misses the rhombus, 
this hypothesis is ruled out, at least on our past light cone. \\

If the latter happens to be the case, we shall need an alternative model 
to fit the data. An example of a model, able to fulfil this purpose 
without the help of a cosmological constant is proposed in the 
following section.

\section{Example: Lema\^itre-Tolman-Bondi model with zero cosmological 
constant}

The complexity of the redshift-distance relation in inhomogeneous 
models and their deviation from the Friedmann relation have recently 
been stressed by Kurki-Suonio and Liang (\cite{kurki-suonio}) and 
Mustapha et al. (\cite{mustapha98}). In a previous very interesting 
paper, Partovi and Mashhoon (\cite{partovi}) have shown that the 
luminosity distance-redshift relation, in local models with radial 
inhomogeneities and the barotropic equation of state for the source 
of gravitational energy, cannot be distinguished from the FLRW one, at 
least to second 
order in $z$. A special case of the solutions studied by these authors 
is used here to explore to which extent such inhomogeneous models 
can mimic homogeneity. \\

A class of spatially spherically symmetrical solutions of Einstein's 
equations, with dust (pressureless ideal gas) as a source of 
gravitational energy, was first proposed by Lema\^itre 
(\cite{lemaitre}). It 
was later on discussed by Tolman (\cite{tolman}) and Bondi 
(\cite{bondi}), and became popular as the ``Tolman-Bondi'' model. 
In the following, it will be refered to as LTB model. \\

Where the Cosmological Principle could to be ruled out, the 
LTB solution 
would appear as a good tool for the study of the observed universe 
in the matter dominated region (\cite{celerier,schneider}). It is 
used here as an example to show that a non vanishing cosmological 
constant in a FLRW universe can be replaced by inhomogeneity 
with a zero cosmological constant to fit the SNIa data. \\

The LTB line-element, in comoving coordinates ($r,\theta,\varphi$) and
proper time $t$, is
\begin{equation}
ds^2=-dt^2 + S^2(r,t)dr^2 + R^2(r,t)(d\theta^2 + \sin^2 \theta 
d\varphi^2) , \label{eq:14}
\end{equation}
in units $c=1$. \\

Einstein's equations, with $\Lambda =$0 and the stress-energy tensor of 
dust, imply the following constraints upon the metric coefficients:
\begin{eqnarray}
S^2(r,t) &=& {R^{'2}(r,t)\over {1+2E(r)}} , \label{eq:15}   \\
{1\over 2} \dot{R}^2(r,t) &-& {GM(r)\over R(r,t)}=E(r) , \label{eq:16} \\
4\pi \rho (r,t) &=& {M'(r) \over R'(r,t) R^2(r,t)} , \label{eq:17}
\end{eqnarray}
where a dot denotes differentiation with respect to $t$ and a prime 
with respect to $r$, and $\rho (r,t)$ is the energy density of the matter. 
$E(r)$ and $M(r)$ are arbitrary functions of $r$. $E(r)$ can be 
interpreted as the total energy per unit mass and $M(r)$ as the mass 
within the sphere of comoving radial coordinate $r$. \\

One easily verifies that Eq.(\ref{eq:16}) possesses solutions for 
$R(r,t)$, which differ owing to the sign of function $E(r)$ and run as 
follows.
\begin{enumerate}
\item with $E(r)>0$, for all $r$
\begin{eqnarray}
R&=&{GM(r)\over 2E(r)} (\cosh u-1) , \label{eq:18} \\
t-t_0(r)&=&{GM(r)\over [2E(r)]^{3/2}}(\sinh u-u) . \nonumber \\
\nonumber
\end{eqnarray}

\item with $E(r)=0$, for all $r$
\begin{eqnarray}
R(r,t)=\left[ {9GM(r)\over 2}\right]^{1/3} [t-t_0(r)]^{2/3} . 
\label{eq:19}
\\ 
\nonumber
\end{eqnarray}

\item with $E(r)<0$, for all $r$
\begin{eqnarray}
R&=&{GM(r)\over -2E(r)}(1-\cos u) ,  \label{eq:20} \\
t-t_0(r)&=&{GM(r)\over [-2E(r)]^{3/2}} (u-\sin u) . \nonumber \\
\nonumber
\end{eqnarray}
\end{enumerate}
where $t_0(r)$ is another arbitrary function of $r$, usually 
interpreted, for 
cosmological use, as a Big-Bang singularity surface \footnote{In fact, 
the physical interpretation of this surface is improper. One can always 
consider that as the energy density increases while reaching its 
neighbourhood, radiation becomes the dominant energy component, 
pressure can no more be neglected, and the LTB model no longer 
holds.}, and for which $R(r,t)=0$. One can choose $t_0(r)=0$ at 
the symmetry center $(r=0)$ by an appropriate translation of the $t=$ 
const. surfaces and describe the universe by the $t>t_0(r)$ part of 
the $(r,t)$ plane, increasing $t$ corresponding to going from the past 
to the future. \\

The physical interpretation of $R(r,t)$ must also be discussed. Bondi 
(\cite{bondi}) presents it as the radial luminosity distance of a 
radiating source. But, with the definition used here (above 
Eq.(\ref{eq:1})) and a corrected expression for apparent luminosity, 
it corresponds in fact to the luminosity distance divided by a factor 
$(1+z)^2$, as is shown below. \\

Assuming that the light wavelength is much smaller than any reasonably 
defined radius of curvature for the universe (geometrical optics 
approximation), Kristian and Sachs (1965) established that the apparent 
intensity of a source, as measured at any point of one of its emitted 
light rays by an observer with proper velocity $u^a$, is
\begin{equation}
l=\mu (k_au^a)^2  , \label{eq:21}
\end{equation}
where $\mu$ is a scalar, corresponding to the magnitude of the 
electromagnetic tensor components as measured by the observer. \\

From the definition of redshift, one obtains (\cite{ellis71})
\begin{equation}
1+z={(k_au^a)_s\over (k_au^a)_o} , \label{eq:22}
\end{equation}
where $k_a$ is tangent to the null geodesics on which light travels, 
the subscripts $s$ and $o$ denoting values at the source 
and at the observer, respectively. \\

For a measure realised at the source by an observer motionless in the 
rest frame of the source
\begin{equation}
(k_au^a)^2_s=1 , \label{eq:23}
\end{equation}
which gives \footnote{Bondi improperly writes: $l=\mu$ ($U=\overline{
E^{\mu \nu}}$ in Bondi's notation).}
\begin{eqnarray}
l={\mu \over {(1+z)^2}} . \label{eq:24}
\\
\nonumber
\end{eqnarray}

Assuming that the electromagnetic tensor for the light emitted by a 
distant source verifies Maxwell's equations for vacuum (\cite{kristian}, 
Bondi \cite{bondi}), it follows, for a radial measurement realised on 
any light ray by an observer located at the symmetry center ($r=0$), 
that
\begin{eqnarray}
l={L\over {4 \pi R^2(1+z)^4}} . \label{eq:25}
\\
\nonumber
\end{eqnarray}

The definition of luminosity distance given by Eq.(\ref{eq:1}) 
yields the expression retained by Partovi and Mashhoon (1984), namely
\begin{equation}
D_L=(1+z)^2R . \label{eq:26}
\end{equation}

One thus sees that the luminosity distance $D_L$ is a function of the 
redshift $z$ and through $R$, of the parameters of the model: $M(r)$, 
$E(r)$ and $t_0(r)$. \\

A light ray issued from a radiating source with coordinates $(t,r,
\theta ,\varphi)$ and radially directed towards an observer located 
at the symmetry center of the model, satisfies, from Eq.(\ref{eq:14}) 
and (\ref{eq:15}),
\begin{eqnarray}
dt=-\,{R'(r,t)\over {\sqrt{1+2E(r)}}}dr . \label{eq:27}
\\
\nonumber
\end{eqnarray}

For a given function $R(r,t)$, Eq.(\ref{eq:27}) possesses an infinite 
number of solutions of the form $t(r)$, depending on the initial 
conditions at the source or on the final conditions at the observer. 
One can thus consider two rays emitted by the same source at slightly 
different times separated by $\tau$. The equation for the first ray 
can be written
\begin{equation}
t=T(r) . \label{eq:28}
\end{equation}

The equation for the second ray is therefore
\begin{equation}
t=T(r) + \tau (r) . \label{eq:29}
\end{equation}

Assuming: $\tau (r) \ll T(r)$ for all $r$, Eq.(\ref{eq:27}) implies
\begin{eqnarray}
{dT(r)\over dr}&=&-\,{R'[r,T(r)]\over \sqrt{1+2 E(r)}} , \label{eq:30} \\
{d\tau (r)\over dr}&=&-\,{\tau (r)\over \sqrt{1+2 E(r)}}
{\partial R'\over \partial t}[r,T(r)]  , \label{eq:31} \\
\nonumber
\end{eqnarray}
which gives the equation of a ray (Eq.(\ref{eq:30})) and the equation 
for the variation of $\tau (r)$ along this ray (Eq.(\ref{eq:31})). \\

If one considers $\tau (r)$ as the period of oscillation of some 
spectral line emitted by the source and $\tau (0)$, its period as 
measured by the observer at $r=0$, the definition of the redshift is
\begin{eqnarray}
\tau (r)={\tau (0)\over {1+z}} . \label{eq:32} \\
\nonumber
\end{eqnarray}

From this equation, another way of writing the rate of variation of 
$\tau (r)$ along a ray is
\begin{eqnarray}
{d\tau (r)\over dr}=-\,\tau (0){\partial z\over \partial r}{1\over 
(1+z)^2} . \label{eq:33} \\
\nonumber
\end{eqnarray}

One can always choose $z$ as a parameter along the null geodesics of 
the rays and obtain from the above equations
\begin{eqnarray}
{dr\over dz}={\sqrt{1+2E(r)}\over {(1+z)\dot {R'}[r,T(r)]}} . 
\label{eq:34} \\
\nonumber
\end{eqnarray}

Eq.(\ref{eq:27}) therefore becomes
\begin{eqnarray}
{dt\over dz}=-\,{R'[r,T(r)]\over {(1+z)\dot {R'}[r,T(r)]}} . 
\label{eq:35} \\
\nonumber
\end{eqnarray}

Eq.(\ref{eq:34}) and (\ref{eq:35}) form a system of two partial 
differential equations of which each null geodesic is a solution 
starting with $z$ at the source and finishing at the observer with 
$z=0$. \\

Successive partial derivatives of $R$ with respect to $r$ and $t$, 
and derivatives of $E(r)$ with respect to $r$, evaluated  at the 
observer, contribute to the expression of the 
coefficients of the luminosity distance (Eq.(\ref{eq:26})) expansion 
in powers of $z$. It is therefore interesting to note 
the behaviour of $R(r,t)$ and $E(r)$ near the symmetry center of 
the model, i.e. near the observer (\cite{humphreys}):
\begin{eqnarray}
R(r,t)&=&R'(0,t)\, r+{\cal O}(r^2) , \label{eq:36} \\
E(r)&=&{1\over 2}E''(0)\, r^2+{\cal O}(r^3) . \label{eq:37} \\
\nonumber
\end{eqnarray}

One thus easily sees that $R$, $\dot{R}$, and higher order derivatives 
of $R$ with respect to $t$ alone, vanish at the observer, as do 
$E$ and $E'$. \\

Expressions for the coefficients of the luminosity distance expansion 
naturally follow; after some calculations, one obtains
\begin{eqnarray}
D^{(1)}_L&=&{R'\over \dot {R'}} , \label{eq:38} \\
D^{(2)}_L&=&{1\over 2}{R'\over \dot {R'}}\left (1+{R'\ddot {R'}\over 
\dot {R'}^2}+{R''\over R'\dot {R'}}-{\dot {R''}\over \dot {R'}^2} 
\right ) , \label{eq:39} \\
D^{(3)}_L&=&{1\over 6}{R'\over \dot {R'}}\biggl(-1-{R'\ddot {R'}\over 
\dot {R'}^2}+3 \, {R'^2\ddot {R'}^2\over \dot {R'}^4}-{R'^2\buildrel 
{...}\over {R'}\over \dot {R'}^3} \nonumber \\
&&
-6 \, {R'\ddot {R'}\dot {R''}\over \dot {R'}^4}
+4 \, {R''\ddot {R'}\over \dot {R'}^3}+2 \, {R'\ddot {R''}\over \dot 
{R'}^3}-3 \, {R''\dot {R''}\over {R'\dot {R'}^3}} \nonumber \\
&&
+3 \, {\dot {R''}^2\over \dot {R'}^4}+{R'''\over R'\dot {R'}^2}
-{\dot {R'''}\over \dot {R'}^3}+{E''\over \dot {R'}^2}\biggr) , 
\label{eq:40} \\
\nonumber
\end{eqnarray}
with implicit evaluation of the partial derivatives at the observer. \\

Following Humphreys, Maartens and Matravers (1997), one can adopt a 
covariant definition for the Hubble and deceleration parameters of 
a spherically symmetric inhomogeneous universe. These authors (see also 
Partovi and Mashhoon (1984)) derive expressions for $H_0$ and $q_0$ at 
the observer, in units $c=1$:
\begin{eqnarray}
H_0={1\over D^{(1)}_L}=\left(\dot {R'}\over R'\right )_0 , 
\label{eq:41} \\
q_0=-H_0 D^{(2)}_L-\, 3 . \label{eq:42} \\
\nonumber
\end{eqnarray}

Substituting into Eq.(\ref{eq:38}) to (\ref{eq:40}), it is easy to see 
that the expressions for the FLRW coefficients in the expansion of 
$D_L$ in powers of $z$ can mimic LTB with $\Lambda=0$  ones, at least 
to third order. This is straightforward for $D^{(1)}_L$. The case of 
$D^{(2)}_L$ is discussed at length in Partovi and Mashhoon 
(\cite{partovi}). For higher order 
terms, it implies constraints on the LTB parameters, which will be 
illustrated below on the peculiar example of flat models. In fact, 
owing to the appearance of higher order derivatives of the parameter 
functions in each higher order coefficient, LTB models 
are completly degenerate with respect to any magnitude-redshift 
relation, while FLRW ones, of which the parameters are constants, 
are more rapidly constrained and cannot thus fit any given relation, 
when tested at sufficiently high redshifts. \\

\section{Illustration: flat LTB ($\Lambda=0$) models}

To illustrate which kind of constraints can be imposed on LTB 
parameters by current observational results, the peculiar case of 
spatially flat LTB ($\Lambda=0$) models is analysed here. \\

Spatial flatness is a property of the subclass of LTB models 
verifying $E(r)=0$ (Bondi 1947). In this case, the expression for $R$ 
is given above by Eq.(\ref{eq:19}) and the calculation of the 
successive derivatives of $R$, contributing to the expressions for 
the expansion coefficients, is straightforward. \\

As the mass function $M(r)$ remains constant with time, it can be used 
to define a radial coordinate $r$: $M(r)\equiv M_0 r^3$, where $M_0$ 
is a constant. \\

With the covariant definition for $H_0$ above mentioned 
(Eq.(\ref{eq:41})), the ${\cal D}^{(i)}_L$s, as derived from 
Eq.(\ref{eq:38}) to (\ref{eq:40}), can thus be written, in units 
$c=1$, in the form
\begin{eqnarray}
D^{(1)}_L&=&{1\over H_0} , \label{eq:43} \\
D^{(2)}_L&=&{1\over 4 H_0}\left (1-6{t'_0(0)\over \left(9GM_0\over 2
\right)^{1\over 3}t_p^{2\over 3}} \right ) , \label{eq:44} \\
D^{(3)}_L&=&{1\over 8 H_0}\biggl(-1+4{t'_0(0)\over \left(9GM_0\over 2
\right)^{1\over 3}t_p^{2\over 3}}+6{t'^2_0(0)\over \left(9GM_0\over 2
\right)^{2\over 3}t_p^{4\over 3}} \nonumber \\
&&-9{t''_0(0)\over \left(9GM_0\over 2
\right)^{2\over 3}t_p^{1\over 3}}\biggr ) , \label{eq:45} \\
\nonumber
\end{eqnarray}
with the previously indicated choice $t_0(0)=0$, where $t_p$ is the 
time-like coordinate at the observer. It is convenient to note 
that $t_p$ is not a free parameter of the model, since its value 
proceeds from the currently measured temperature at 2.73 K 
(\cite{celerier}). \\

A comparison with the corresponding FLRW coefficients gives the 
following relations:
\begin{eqnarray}
\Omega _M &\longleftrightarrow& 1+5\, {t'_0(0)\over \left(9GM_0\over 
2\right)^{1\over 3}t_p^{2\over 3}}+{15\over 2}\, 
{t'^2_0(0)\over \left(9GM_0\over 2\right)^{2\over 3}t_p^{4\over 3}}
 \nonumber \\
&&+{9\over 4}\, {t''_0(0)\over \left(9GM_0\over 2\right)^{2\over 3}
t_p^{1\over 3}} , \label{eq:46} \\
\Omega _\Lambda &\longleftrightarrow& -\, {1\over 2}\, {t'_0(0)\over 
\left(9GM_0\over 2\right)^{1\over 3}t_p^{2\over 3}}+{15\over 4}\, 
{t'^2_0(0)\over \left(9GM_0\over 2\right)^{2\over 3}
t_p^{4\over 3}} \nonumber \\
&&+{9\over 8}\, {t''_0(0)\over \left(9GM_0\over 
2\right)^{2\over 3}t_p^{1\over 3}} . \label{eq:47} \\
\nonumber
\end{eqnarray}
 
The above Eq.(\ref{eq:47}) implies that a 
non vanishing cosmological constant in a FLRW interpretation of data 
at $z<1$ corresponds to a mere constraint on the model parameters in 
a flat LTB ($\Lambda =0$) interpretation. \\

Any magnitude-redshift relation, established up to the redshifts and 
with the precisions achieved by current measurements, i.e. at third 
order level, can thus be 
interpreted in either model. For instance, the latest results 
published in P99, and given under the form of Eq.(\ref{eq:11}), 
correspond, in a flat LTB ($\Lambda =0$) interpretation, to
\begin{eqnarray}
&&4.3\, {t'_0(0)\over \left(9GM_0\over 2\right)^{1\over 3}t_p^{2\over 3}}
+3.75\, {t'^2_0(0)\over \left(9GM_0\over 2\right)^{2\over 3}t_p^{4\over 
3}}+1.125\, {t''_0(0)\over \left(9GM_0\over 2\right)^{2\over 3}
t_p^{1\over 3}} \nonumber \\
&&\approx -1\, \pm \, 0.1 . \label{eq:48}
\end{eqnarray}

Such a result would imply a negative value for at least one of the 
two quantities $t'_0(0)$ or $t''_0(0)$, which would be an interesting 
constraint on the ``Big-Bang'' function in the observer's 
neighbourhood. For instance, a function $t_0(r)$ decreasing near the 
observer would imply, for a source at a given $z<1$, an elapsed time 
from the initial singularity that is longer in an LTB model than in 
the corresponding FLRW one, i.e. an ``older'' universe \footnote{In 
fact, see footnote 3, a lower temperature.}. A decreasing $t_0(r)$ 
has thus an analogous effect in a LTB universe to a positive 
cosmological constant in a FLRW one. They both make the observed 
universe look ``older''.

\section{Conclusion and discussion}

The interpretation of recently published data from high redshift SNIa 
surveys, devoted to the measurement of the magnitude-redshift 
relation, has been 
re-examined with no a priori idea about which model would best describe 
our universe, at least up to redshifts $z\sim 1$. \\

It has been shown that a straight reading of these data does 
not exclude the possibility of ruling out the Cosmological Principle. 
\\

A method of testing large scale homogeneity on our past light 
cone, provided sufficiently accurate data from ``standard 
candles'' at redshifts approaching $z=1$ would be available, has been 
proposed. This method could be applied using the SNIa data, provided 
every source of potential bias or systematic uncertainties would be 
correctly taken into account. An evolution of the progenitors is 
generally considered as one of the 
more likely sources of potential systematic errors in the analysis 
of the supernovae data. Mustapha et al. (\cite{mustapha97}) have made 
the point that, given isotropic observations about us, homogeneity 
cannot be proved without either a model independent theory of source 
evolution, or distance measures that are independent of source 
evolution. If a perceptible evolution effect was to be put forward, 
its impact on the SNIa data would thus have to be evaluated with a 
model independent method, before using them as a test for 
homogeneity. \\

An example of an inhomogeneous model with a zero cosmological constant 
reproducing the current observations has been given. However, a 
vanishing cosmological constant is a minimal feature here imposed on 
the model only to prove that $\Lambda $ is not necessary to fulfill 
the observational constraints. A non zero $\Lambda$ would only add a 
new free parameter in the LTB equations and would not alter the 
following primary statement: the presently 
published SNIa data can be interpreted as implying either a 
strictly positive cosmological constant in an homogeneous universe or 
large scale inhomogeneity with no constraint on $\Lambda$. \\

The choice of over-simple inhomogeneous models as examples can, of 
course, be questioned. The most naive assumptions of spherical 
symmetry and centered observer, which have been discussed 
in other works (see e.g. \cite{celerier}, \cite{celerier99}), have been 
mainly retained for simplification purpose, but do not distort the 
above conclusions. It could be actually interesting to probe the 
possibility of an off-center observer (see e.g. \cite{humphreys}) 
with the SNIa data. Spherical symmetry is grounded on the 
observed quasi-anisotropy of the CMBR temperature, and can thus be 
considered as a sufficiently good working approximation. \\

The possibility of testing finer features of the model 
depends on observational improvements. But before aiming at such 
ambitious goals, it is of the utmost importance to complete the large 
scale homogeneity test. A previous increase in the number and 
measurement accuracy of the candidate ``standard candles'' at very high 
redshifts is therefore urgently needed, for progress in both 
fundamental issues of the Cosmological Principle and of the 
cosmological constant. \\

It has been shown however that LTB models are highly degenerate 
with respect to any magnitude-redshift relation, but this is not the 
case for FLRW models. The best way to prove large scale inhomogeneity 
would therefore be to disprove homogeneity. Conversely, the best 
way to prove a non zero cosmological constant would be to prove large 
scale homogeneity. But, as the Friedmann distance-redshift relation is 
a necessary but not a sufficient condition for homogeneity, its 
observed verification would not, in principle, be enough to support the 
Cosmological Principle. Even if this would imply a fine tuning of its 
parameters, the possibility for an inhomogeneous universe to mimic such 
a relation could not be excluded. \\

To consolidate the robustness of the future magnitude-redshift tests, it 
would therefore be worth confronting their results with the full range 
of available cosmological data. The most recent attempt to do so is the 
work by Bahcall et al. (\cite{bahcall}) who conclude that the analysed 
observations can be considered as consistent with the standard Big-Bang 
picture of the expansion of the universe, but do not discuss the 
homogeneity hypothesis. A cross-check with a model independent analysis 
of the CMBR anisotropy would, for instance, likely yield a substantial 
improvement, as the CMBR data are known to provide orthogonal 
constraints on the models parameters. \\

{\it Aknowledgements.} The author thanks Brandon Carter for his 
help in correcting the manuscript.

\end{document}